# The decrease of the critical current of coated conductors when a perpendicular magnetic field is applied : a Josephson effect point of view

C. McLoughlin[a,d], Y. Thimont[a,b,c], J. Noudem[a,b,c], C. Harnois[a,b,c] and P. Bernstein[a,b,c]

[a]CRISMAT-ENSICAEN, UMR-CNRS 6508 F14050 Caen, France
[b]Université de Caen-Basse Normandie, Caen, France
[c]CNRS, UMR6508, Caen France
[d]National Centre for Plasma Science and Technology, School of Physical Sciences, Dublin City University, Glasnevin, Dublin 9, Ireland

**Abstract**

A large decrease is observed in the critical current density of YBCO coated conductors (CC) and related compounds when a strong perpendicular magnetic field is applied. While measurements are generally carried out at 77K only, here we present a magnetic technique permitting to determine the critical current per unit width of conductor ($I_{cr}/w$) in a large temperature range. We report measurements carried out on various CCs that show that, in addition to the reduction in the critical temperature that can be attributed to the low value of the irreversibility field near $T_c$, the field application results in a large decrease of $I_{cr}/w$ at all temperatures. We ascribe this reduction to the Josephson behaviour of the twin boundaries included in the YBCO layer.

## 1. Introduction

In many applications considered for coated conductors (CC) such as motors, generators or superconducting magnetic energy storage systems (SMES), the CCs must keep a large critical current $I_{cr}$, in presence of a strong magnetic field B. For second generation CCs, whose superconducting layer belongs to the YBCO family, most studies on this question consist of the measurement of the critical current as a function of the amplitude and of the direction of the applied magnetic field at 77K or a few other temperatures (see for examples [1,2,3]). They show a strong $I_{cr}$ anisotropy, the decrease in the critical current being much larger for a perpendicular than for an in-plane field. In order to limit the $I_{cr}$ reduction in the perpendicular configuration, a common practice is to reinforce the pinning properties of the materials, generally by doping the superconducting layer with $BaZrO_3$ nano-rods or a similar material. Very few studies, however, report the behaviour of $I_{cr}(B)$ in a large range of temperatures. The main reason is that, due to the large amplitude of the critical current of the CCs at low temperature, these measurements are complicated and time-consuming with conventional methods.

In this contribution, in section 2, we describe a technique permitting to determine the critical current per unit width of conductor ($I_{cr}/w$) in the whole range of the temperatures from magnetic measurements. In section 3, we report the $I_{cr}/w$ values of various commercial CCs between 10K and their critical temperature, $T_c$, when a perpendicular field with an amplitude ranging between 0 and 5T is applied. The results show that, in addition to the expected reduction of the temperature at which the critical current is non-zero, there is a strong reduction of $I_{cr}/w$ at all temperatures. We attribute this reduction in section 4 to the Josephson behaviour of the twin boundaries included in the superconducting layer.

**2. Determining $I_{cr}/w$ from magnetic measurements**

*2.1 Measurements with no applied field*

Samples with dimensions 4x4 mm$^2$ were cut in the investigated CCs and cooled down to 10K in a cryostat. At this temperature, a $H_a$=5T field was applied perpendicular to the sample plane and subsequently switched off. The same procedure was carried out with a reverse field. This procedure aims at erasing the magnetic history of the sample and to generate screening currents in the CC superconducting layer. Then, the magnetic moment of the sample, m(T), that is due to the currents persisting in the superconducting film after the suppression of the applied field, was measured via a SQUID magnetometer while increasing the temperature between 10K and 100K. From classical electro-magnetism, the magnetic moment resulting from the currents circulating in a film with surface S takes the form :

$$\boldsymbol{m} = \frac{1}{2}\iint_S \boldsymbol{r} \times \boldsymbol{J}^S dS \qquad (1)$$

where $J^S$ is the sheet current density or current per unit width. For a large enough field, the sheet density of the current induced by the application of the field is equal to the critical value and we can write $J^S = I_{cr}/w$ everywhere in the film. Here, we'll consider that the persisting current lines in square samples are circular, as shown in Fig.1. This hypothesis is different from that we previously made in [4] but, from measurements carried out on numerous samples, it has turned out that it yields more accurate results. From Eq.(1) and the symmetry of the current lines, the critical current per unit width of conductor can be written as :

$$\frac{I_{cr}}{w} = -\frac{24m}{\pi w^3 \left[1 + \frac{(\sqrt{2}-1)^3}{2}\right]} \qquad (2)$$

*2.2 Measurements with an applied field.*
The measurement procedure is the same as that described in section 2.1, except that the measuring field is applied during the measurements. For fields large enough for the screening current density to be equal to the critical value, the relation between $I_{cr}/w$ and the measured magnetic moment is also given by Eq.(2) .

**3. Results**

Fig.2, 3 and 4 show the $I_{cr}/w$ obtained for two CCs made by SuperPower and one made by American Superconductor. We stress that the values reported for this last CC when, a field is applied are less accurate than for the Superpower ones, because of the contribution to the measured magnetic moment of its strongly magnetic substrate. We have subtracted this contribution, but this procedure has resulted in an increase in the error on $I_{cr}/w$. The validity of the technique we have used to determine $I_{cr}/w$ was confirmed by transport measurements with no applied field carried out on the Superpower SCS4050 by Th.Lécrevisse *et al.* [5]. As seen in Fig.2, their measurements fall on the curve obtained from the magnetic measurements. The field application results in two phenomena for all the samples : i) a reduction in the critical temperature, increasing with the field amplitude and ii) a reduction in $I_{cr}/w$ in the whole range of the temperatures. We observe in addition that the curves measured with B=5T are very similar for the three investigated CCs. The reduced critical temperature, $T_c(B)$, can probably to be ascribed to the low value of the irreversibility field near $T_c(0)$ as compared to that of the applied field. The CCs are still in the superconducting state between $T_c(B)$ and $T_c(0)$ but, since the vortices flow freely in the superconducting layer, the critical current is zero. We suggest now that, in a large temperature range, the reduction in $I_{cr}/w$ can be attributed to the Josephson behaviour of the twin boundaries (TB) included in the superconducting layer.

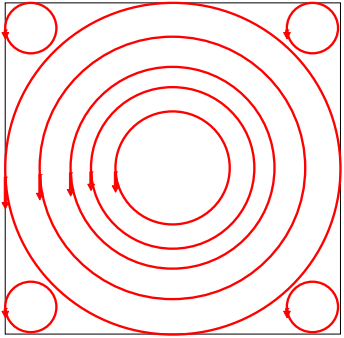 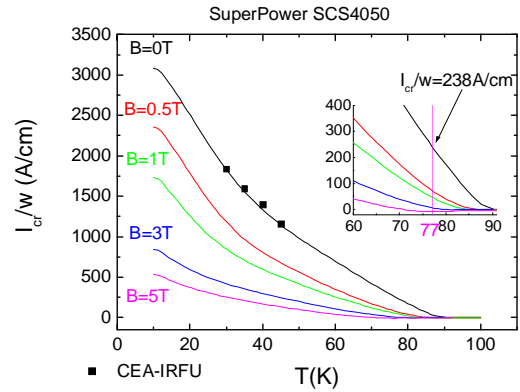

Fig.1 : Assumed shape for the persisting current lines flowing in the samples.

Fig.2 : Critical current per centimeter width measured on the SuperPower SCS4050 sample as a function of the temperature with various applied fields. The inset shows the same curves in the high temperature range. The full squares are transport measurements carried out on the same sample by Th.Lécrevisse of CEA-IRFU [5]

## 4. Model

*4.1 Critical current per unit width when no field is applied*

YBCO films and related compounds include twin boundaries (TB) that play an important role in their transport properties. Since their width is in the range of $\xi_{ab}(T)$, the superconducting coherence length in the a-b planes of the material, a tunneling pair current flowing across the TBs and the modulation of the critical current by a magnetic field can be expected. However, another feature of the TBs is that they are the preferred location for defects with a size comparable to or larger than $\xi_{ab}(T)$. These defects can be large enough to cause locally a disruption of the tunneling current. In previous works, we have assumed that, as a consequence, the boundary planes split off into rows of Josephson weak links (JWL) whose length depends on the temperature [6]. Otherwise, detailed measurements of the temperature dependence of $I_{cr}/w$ carried out on epitaxial YBCO films has suggested that their superconducting critical temperature is modulated along the c-axis [4] and that the TBs can be described as $\nu$ groups of $Z_k$ rows of JWLs with the same critical temperature $T_{ck}$. Considering an individual JWL, it is reasonable to expect that its current amplitude, $I_J(0)$, is equal to the minimum current maintaining the phase coherence against thermal fluctuations. The corresponding JWL energy is equal to $k_B T$ and we have :

$$I_J(0) = \frac{2\pi k_B T}{\phi_0} . \qquad (3).$$

In the group of $Z_k$ rows with critical temperature $T_{ck}$, the current carried by a row of JWLs takes the form :

$$I_{cr_k} = I_J(0) N_k \qquad (4)$$

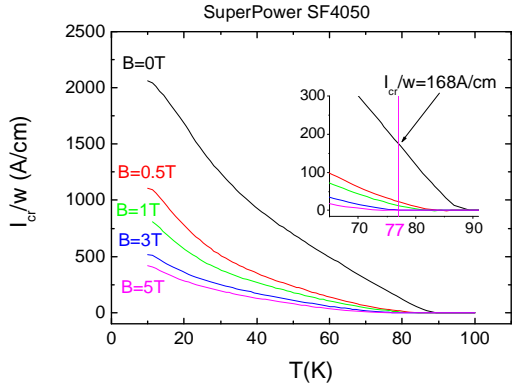 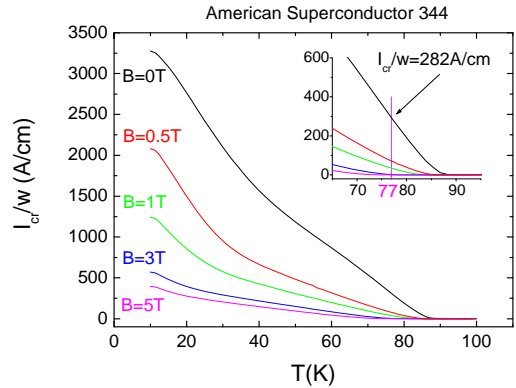

Fig.3 : Critical current per centimeter width measured on the SuperPower SF4050 sample as a function of the temperature with various applied fields. The inset shows the same curves in the high temperature range.

Fig.4 : Critical current per centimeter width measured on the American Superconductor 344 sample as a function of the temperature with various applied fields. The inset shows the same curves in the high temperature range.

where $N_k$ is the mean number of weak links in a row that can be written as :

$$N_k = \frac{w}{\langle \delta_k \rangle} \qquad (5).$$

Here $\langle \delta_k \rangle$ is the mean length of the JWLs along the TBs. In previous works [4], we have established that a good approximation for $\langle \delta_k \rangle$ is

$$\langle \delta_k \rangle = \delta_0 \left(1 - \left[T/T_{ck}\right]\right)^{-\frac{3}{2}} \text{ with } \delta_0 = 0.55 nm \qquad (6)$$

Taking into account the contribution of each group of weak links rows, the sample critical current per unit width takes the form

$$\frac{I_{cr}}{w} = \sum_k Z_k \frac{I_{cr\,k}}{w} \qquad (7).$$

With Eqs.(3-7) we can reproduce the $I_{cr}/w$ curves of a sample if we can determine its critical temperature profile, i.e. its $Z_k$ and $T_{ck}$. As an example, Table I reports the $Z_k$ and $T_{ck}$ values fitting the zero field curve of sample SCS4050 in Fig.5. In the same figure we have reported the measured and fitted zero field curves of samples SF4050 and 344. The very good agreement between the experimental and the fitted curves show that we have determined for each sample the correct sets of $T_{ck}$ and $Z_k$ values.

Table 1 : Critical temperature profile of sample SCS4050

| $T_{ck}$ (K) | 90.3 | 69 | 58,5 | 51 | 46,5 | 40 | 34.6 | 30.8 | 27 | 23.5 | 20.4 | 17.3 |
|---|---|---|---|---|---|---|---|---|---|---|---|---|
| $Z_k$ | 75 | 15 | 20 | 30 | 35 | 60 | 40 | 45 | 55 | 50 | 45 | 60 |

*4.2 Critical current per unit width when a field is applied*
When a field is applied, the critical current in JWL p with critical temperature $T_{ck}$ takes the form

$$I_k^p(B) = I_J(0) \left| \frac{\sin\left(\pi \Phi_k^p / \phi_o + \Psi_k^p\right)}{\pi \Phi_k^p / \phi_o} \right| \qquad (8)$$

Here $\Phi_k^p$ is the magnetic flux across the JWL. We can write :

$$\Phi_k^p \cong \Phi_k = 2\lambda_{ab} \langle \delta_k \rangle B \qquad (9)$$

where

$$\lambda_{ab}(T/T_{ck}) = \lambda_{ab0} \left(1 - (T/T_{ck})^2\right)^{-\frac{1}{2}} \qquad (10)$$

is the superconducting penetration length. The term $\psi^p{}_k$ is due to the presence of vortices in the TBs, that results in the absence of coherence in the phase of the order parameter between two JWLs located in the same row. Considering all the JWLs of the row, $\psi^p{}_k$ can be regarded as taking random values. Then, the critical current of row k can be written as :

$$I_{cr_k} = \sum_{p=1}^{N_k} I_k^p = \frac{N_k I_J(0)}{\frac{\pi \Phi_k}{\Phi_0}} \left\langle \left| \sin\left(\frac{\pi \Phi_k}{\phi_o} + \Psi_k^p\right) \right| \right\rangle = \frac{N_k I_J(0)}{\frac{\pi \Phi_k}{\Phi_0}} \times \frac{2}{\pi} \qquad (11)$$

if $\Phi_k$ is large enough for $\psi^p_k$ to take any value between 0 and $\pi/2$. As a consequence of this restriction we'll be able to fit only the curves measured at high field. Otherwise, it is reasonable to expect that no tunneling current flows across the JWLs carrying vortex cores. As a result, $N_k$ in Eq.(11) takes the form :

$$N_k = \frac{w}{\langle \delta_k \rangle} - \frac{w}{d_v} \qquad (12)$$

where $d_v = \sqrt{\phi_0/B}$ is the mean inter-vortex distance. Finally, $I_{cr}/w$ (B) can be calculated with Eq.(7 and 9-12). The values calculated for the measurements carried out at B=5T with the same $T_{ck}$ and $Z_k$ as for $I_{cr}/w$ (0) are reported in Fig.5 taking $\lambda_{ab}(0)=0.19\mu m$. There is a good agreement between calculations and measurements for the SuperPower samples, but some discrepancy is observed at low temperature for the American Superconductor one. This is probably linked to the corrections that were carried out to take into account the magnetic contribution of the substrate to the measurements, as explained above. The fitted curves show a reduction in the critical temperature resulting from the form of $N_k$ in Eq.(12), but this reduction is always smaller than that observed experimentally. This last one, as detailed above, is probably due to the fusion of the vortex lattice near $T_c$ when a large enough field is applied. Ultimately however, the results reported here suggest that it is reasonable to ascribe the $I_{cr}/w$ decrease at temperatures far from $T_c$ to the Josephson behaviour of the twin boundaries.

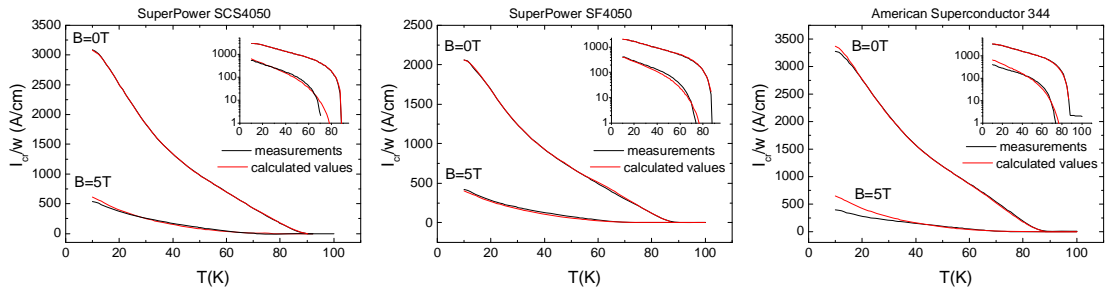

Fig.5 : Experimental and calculated $I_{cr}/w$ values at B=0 and B=5T for the investigated coated conductors. The insets show the same curves on a semi-log scale.